\documentclass[journal]{IEEEtran}

\usepackage{color,soul}
\usepackage{cite}
\usepackage{amsfonts}

\usepackage{amsthm}

\usepackage{multirow}
\usepackage{footnote}
\usepackage{xcolor}
\usepackage{colortbl}
\usepackage{gensymb}
\usepackage{stfloats}
\usepackage{url}
\usepackage{subfigure}
\usepackage{balance}

\usepackage{lipsum} 
\usepackage{fancyhdr}
\ifCLASSINFOpdf
  \usepackage[pdftex]{graphicx}
  \graphicspath{{../pdf/}{../jpeg/}}
  \DeclareGraphicsExtensions{.pdf,.jpeg,.png}
\else
\fi

\usepackage{amsmath}
\usepackage{algorithmic}

\ifCLASSOPTIONcompsoc
  \usepackage[caption=false,font=normalsize,labelfont=sf,textfont=sf, subrefformat=parens,labelformat=parens]{subfig}
\else
  \usepackage[caption=false,font=footnotesize, subrefformat=parens,labelformat=parens]{subfig}
\fi

\begin{document}

\title{Meeting the traffic requirements of residential users in the next decade with current FTTH standards: how much? how long?}

\author{\IEEEauthorblockN{Jos\'{e} Alberto Hern\'{a}ndez, Rafael S\'{a}nchez, Ignacio Mart\'{i}n and David Larrabeiti}\\
\thanks{The authors are with the Department of Telematics Engineering, Universidad Carlos III de Madrid, Spain. }
\IEEEauthorblockA{Universidad Carlos III de Madrid, Spain\\
Email: \{jahgutie, rsfuente, ignmarti, dlarra\}@it.uc3m.es}}



\IEEEtitleabstractindextext{
\begin{abstract}

Traffic demand in the access has grown in the last years, and service providers need to upgrade their infrastructure to the latest access standards. While fiber has become the preferred technology of choice in access networks, there are many fibre access technologies available in the market. This poses a challenging question to operators not always easy to answer: how to upgrade? what technology and for how long it will cope with the demands? In this paper we model the traffic forecast in the access for the next decade and analyze possible upgrade scenarios of fibre access networks, concluding which of the NG-PON flavors could better fit the demand.

\end{abstract}

}

\maketitle

\IEEEdisplaynontitleabstractindextext

\IEEEpeerreviewmaketitle

\section{Introduction}
\label{sec:introduction}

According to~\cite{analysysmason2}, it is estimated that about 90\% of the households in Western Europe will have "superfast broadband connection"  by year 2020. Such a "superfast broadband connection" includes all those technologies capable of delivering download speeds over 30~Mb/s.

To satisfy the ever-increasing demands of users traffic, estimated to grow about 24\% per year~\cite{cisco}, network operators have a wide range of high-speed access technologies to choose from, namely Passive Optical Networks (PONs), Hybrid Fibre-Coax (HFC), cable-based like Digital Subscriber Line (DSL) fast versions like VDSL2 and G.fast\footnote{ITU-T Rec. G.9701 Fast access to subscriber terminals (G.fast), approved in 2014}, the latter capable of providing up to 2~Gb/s. In addition, fixed wireless loop (for instance based on millimeter Wave technology) can also be used in combination with fiber/cable~\cite{rokkas} or in hardly accessible geographical locations. 

In the past decade, a large number of telecommunications operators (aka telcos) have widely deployed Fibre-To-The-x (FTTx) technologies, specially in large cities, where fiber was taken directly to the end user as in Fibre to the Home (FTTH) or very close to him/her as in Fibre to the Neighbourhood/Curb (FTTN/C) and then terminated with cable. 

A PON architecture is based on a shared point-to-multipoint architecture with one or two wavelengths in the downstream direction (from Central Office to users) and one wavelength in the upstream (from users to Central Office). TDM-PON uses a 1:N passive splitter/combiner to divide the optical signal to all users in the downstream direction and aggregate the users'  data in the upstream direction. The OLT uses a Dynamic Bandwidth Allocation (DBA) algorithm to arbitrate access to the shared channel in the upstream direction, avoid collisions, assign bandwidth to the users and provide Quality of Service (QoS) to different types of flows.

At present, some network operators have begun to offer 1 Gb/s Internet access, shared by a number of residential households, typically 32 or 64 in a tree PON topology.  In theory, 64 users sharing a 1~Gb/s PON tree would only receive about 15~Mb/s on average each; however users are typically promised higher bandwidth capacities than that (50, 100 Mb/s and even 1~Gb/s in many countries). This is possible since network operators have realized that their PONs are underutilized. This fact allows operators to employ over-subscription strategies~\cite{rafa_commag}, leveraging the statistical multiplexing gains to reduce the cost of deployment.

Recently, new PON standards have been approved by both the \emph{ITU-T} and the \emph{IEEE} to increase towards 10 Gb/s speed and beyond: XG-PON\footnote{ITU-T Rec. G.987 10-Gigabit-capable passive optical network (XG-PON) systems, approved in 2012} operating at 10G/2.5G DS/US (i.e. Downstream/Upstream), symmetrical XGS-PON\footnote{ITU-T Rec. G.9807 10-Gigabit-capable symmetric passive optical network (XGS-PON), approved in 2016} operating at 10G/10G DS/US, 10G-EPON\footnote{IEEE 802.3av 10~Gb/s Ethernet Passive Optical Network, approved in 2009} and NG-PON2\footnote{ITU-T Rec. G.989 40-Gigabit-capable passive optical networks 2 (NG-PON2), approved in 2014}. This latter standard has already targeted bandwidth capacity beyond 10 Gb/s in PONs currently offering 4 downlink and 4 uplink wavelengths   operating at 10/2.5G each. Finally, the 100G-EPON Task Force\footnote{IEEE P802.3ca 100G-EPON Task Force, expected by 2020} aims to enable 4 carriers at 25G by year 2020.

With so many PON technologies to choose from, network operators should carefully analyse how to design their migration strategy towards the deployment of the new Next-Generation Optical Access (NGOA) technologies, considering the expected traffic demand over the next decade.  Convergence within the same infrastructure for services requiring different service levels should also be considered. Residential, business and mobile traffic are expected to be transported over the same Optical Distribution Network (ODN) infrastucture~\cite{Pon_roadmap}. Operators have a wide range of options to choose from:

\begin{itemize}
\item Deploy new PONs with small split ratios, thus allowing more capacity per household.
\item Upgrade to new standards with more capacity while keeping existing reach and sharing ratios. 
\item Upgrade to new technologies that both provide more capacity but also allows increased reach and sharing ratios. 
\end{itemize}

Can those three approaches cover the traffic demand in the short-, medium- and long-term, respectively? How long can a service provider wait before migrating? This article attempts to answer those questions and shed light into what features are expected to be more demanded from NGOAs in the next years. 

Thus, the rest of the paper is organised as follows: Section II provides a traffic forecast for residential users in the next decade. Section III outlines a traffic model of residential households in a PON architecture. Section IV covers which PON technologies better fit taking into account the expected traffic demand in the next decade. Section V analyzes evolution paths for the next years, when existing PON technologies can not cover traffic demand, also paying attention at the economic aspects of the different migration strategies. Finally, Section VI concludes this article with a summary of its main results and conclusions.

\section{Traffic forecast per residential household in the next decade}

At present, the average traffic generated per residential user in a PON has been observed to span only a few hundred of Kb/s per user, showing that fixed access networks are typically underutilized. Indeed, according to Cisco's estimates for Western Europe, the average traffic generated per person and month will grow from 27.2 GB in 2015 to 66.5 GB in 2020~\cite{cisco}.


In this work, we shall use the numbers provided by the CNMC\footnote{"Comision Nacional de los Mercados y la Competencia", dependent on the Ministry of Economy and Competitiveness of Spain} white paper~\cite{cnmc}, which are estimates obtained from real measurements in fixed access networks of Spain. In this report, the average traffic consumed per Spanish household in 2016 is 77.66 GB per month (see page 94 of~\cite{cnmc}). This number translates to an average bandwidth consumption of approximately 236~Kb/s.

\begin{table}[!htbp]
\centering
\begin{tabular}{| r | c |  c | c | c | c |}
\hline
   & 2016 & 2020 & 2025 & 2030 & 2035 \\
\hline
Avg. Bandwidth (Mb/s)   & 0.236 & 0.577 & 1.76 & 5.38 &16.4 \\
Peak as 3xAvg (Mb/s)  & 0.709 & 1.73  & 5.28 & 16.1 & 49.2 \\
Peak as 5xAvg   (Mb/s) & 1.18 &  2.88 & 8.80 & 26.9 & 82.0 \\
\hline
\end{tabular}
\caption{Residential traffic estimates per household for the next decade. CAGR = 25\%}
\label{tab:trafficforecasts}
\end{table}

In addition, it is worth remarking that peak traffic demands at households typically occur during evenings showing peak-to-average values ranging from 3x to 5x~\cite{analysysmason1}, i.e. between 709~Kb/s and 1.18~Mb/s per household at year 2016. Table~\ref{tab:trafficforecasts} translates these numbers to bandwidth demand estimates for the next decades, assuming a Compound Annual Grouth Rate (CAGR) of 25\% in line with the estimates of the Cisco's Visual Networking Index (see Table 1 of~\cite{cisco}). As shown, the numbers of traffic estimates per residential household by year 2035 reaches 16.4~Mb/s per household on average, and almost 82~Mb/s at peak hours.

Finally, it is also worth noticing that the above numbers are average values per household. However, it has been observed in many scenarios some sort of Pareto-like behavior, where a few number of users generate most of the traffic (aka \emph{heavy hitters}), while the vast majority of users generate only a small portion of the total share. This is further illustrated in the next section and modeled using a Zipf distribution. 

\section{Residential traffic characterization and model}

\begin{figure}[htbp]
\centering
\includegraphics[width=\columnwidth]{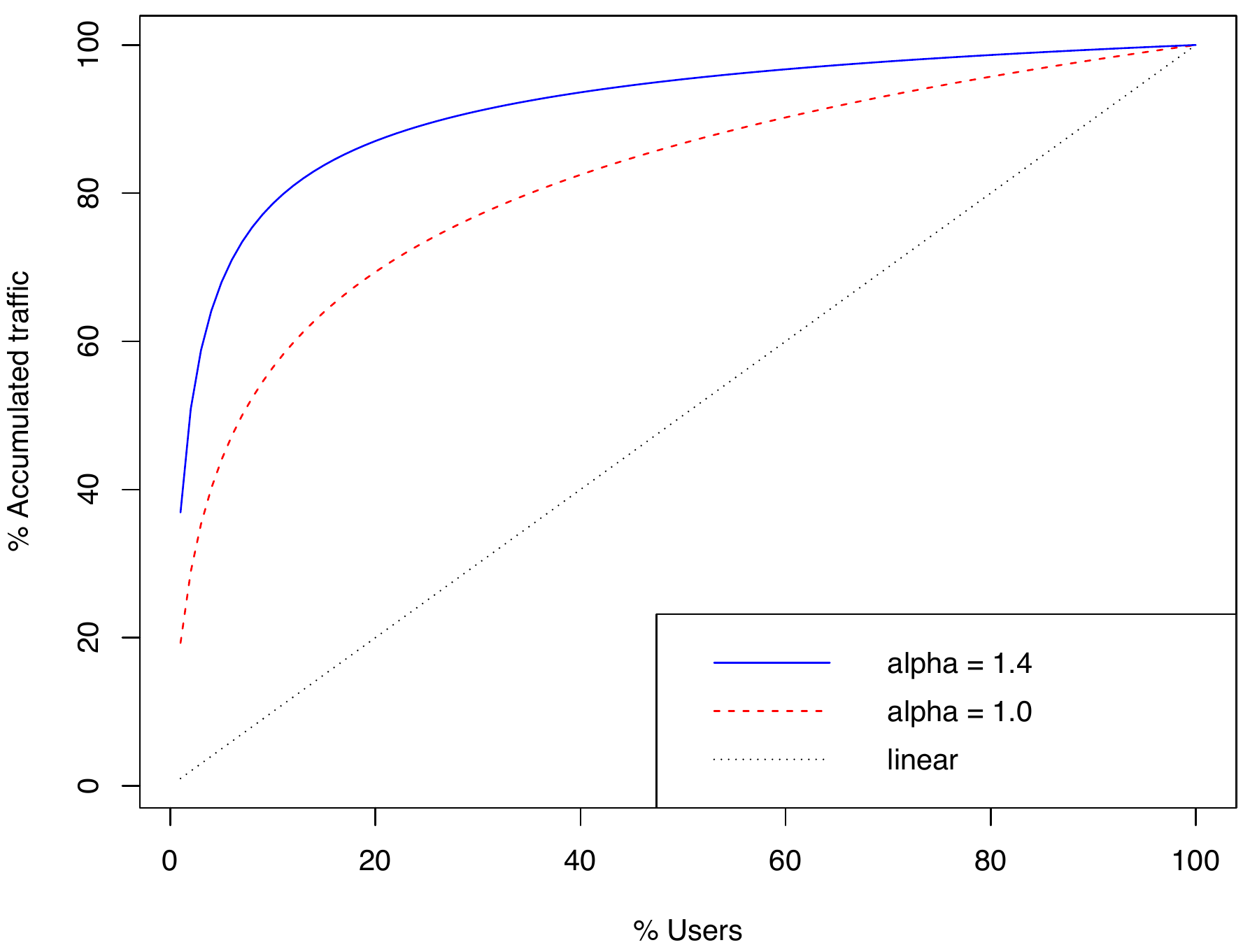}
\caption{Zipf-like traffic profiles (CDF)}
\label{fig:zipf}
\end{figure}

Consider a PON with $N$ users (i.e. split 1:N), each one offering an amount of traffic $B_i$ modeled by a random variable following a Zipf distribution characterized by $N$ and shape parameter $\alpha$. Fig.~\ref{fig:zipf} shows an example of the Cummulative Distribution Function (CDF) of two Zipf distributions with $N=100$ users and shape parameters $\alpha=1.0$ and $1.4$; in addition, the figure shows a linear example case where all users generate the same average bandwidth. 

 In the case of shape parameter $\alpha = 1.0$, Fig.~\ref{fig:zipf} reveals that the top-3\% of users generate 35\% of the total traffic share, while the top-10\% of the users generate 56\% of the total traffic share. For $\alpha = 1.4$, the shares are even more disparate, namely the top-3\% of users generate 59\% of the total traffic share, while the top-10\% of the users generate 78\% of the total traffic share.Thus, shape parameter $\alpha$ determines how uneven users behave regarding traffic generation: the larger its value, the more user traffic variability.

In the following experiments, we shall use the shape parameter $\alpha = 1.0$, in line with the Ericsson mobility report~\cite{ericsson2015}, which clusters users into five categories: light, medium, medium-high, heavy and extreme users, showing that only 10\% of the users generate 55\% of the data traffic (see~\cite{ericsson2015}, page 24). 

According to this model, the top-3 heaviest users offer 22.8, 11.4 and 7.6~Mb/s respectively, while the top-3 lightest users offer only 0.232, 0.230 and 0.228~Mb/s respectively. The user average is 1.18~Mb/s and the standard deviation is 2.68~Mb/s. These numbers are in line with Table~\ref{tab:trafficforecasts}, Peak as 5xAvg, year 2016.

\section{Meeting the traffic needs of residential users in the next decade}

Now, let $B_N^{tot}$ define the total traffic offered by the $N$ users sharing a PON with split 1:N, i.e. $B_N^{tot} = B_1 + \ldots + B_N$, where the $B_i$ are independent and identically distributed (iid) random variables following the Zipf distribution overviewed in the previous section. Typical numbers for $N$ go from split ratios of 1:8 to 1:64 for most typical PONs, and is expected to grow to 1:128 and even 1:256 in the future as specified in the standards. For the sake of theoretical completeness, let us also consider very-high logical splits up to 1:1024 and long reach (up to 125 km), as envisioned by the case of Long-Range PONs (LR-PONs)~\cite{ruffini}. It is worth remarking that, at present, LR-PONs are not a standard, only purely scientific work.

\begin{figure}[htbp]
\centering
\includegraphics[width=\columnwidth]{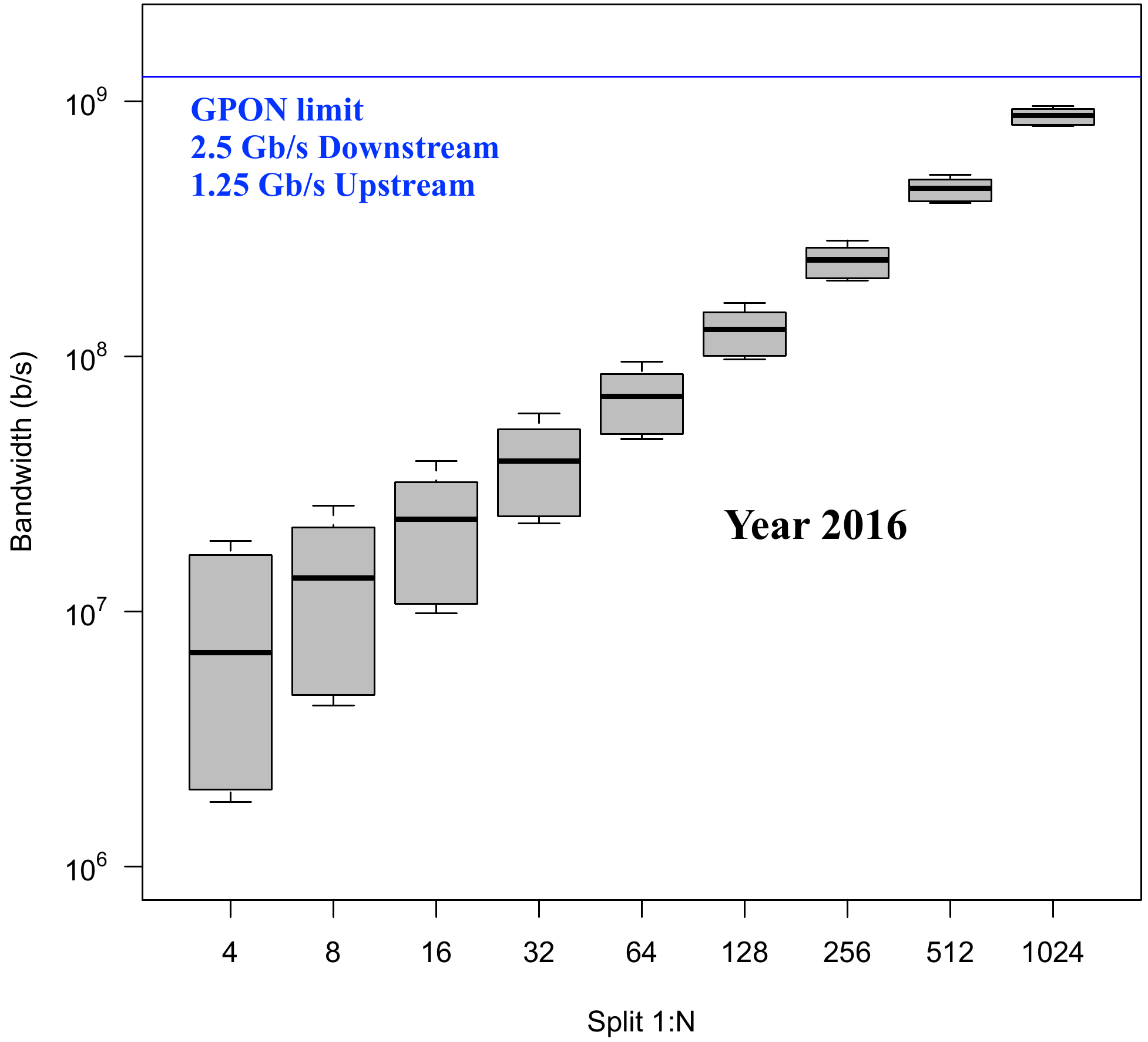}
\caption{Aggregated offered traffic per 1:N PON (Year 2016)}
\label{fig:year2017}
\end{figure}

Fig.~\ref{fig:year2017} shows some boxplots summarizing the simulation results for the traffic needs of PONs at year 2016, i.e. peak traffic of 1.18~Mb/s per household. The simulations have been conducted using the Bootstrap method, which relies on random sampling with replacement to obtain accurate estimates of parameters (see~\cite{efron_bootstrap} for further details). 

The figure covers PONs with split ratios ranging from 1:4 up to 1:1024. The horizontal blue line limit represents the 1.25~Gb/s capacity limit in the upstream direction of GPON. This figure reveals the following interesting observations:
\begin{itemize}
\item Low split ratios (i.e. 1:4, 1:8) show high variability in terms of aggregated offered traffic. Essentially, when designing a PON for 4 users, it may happen that such four users sharing the PON are all either heavy hitters or light users; thus one may find 1:4 PONs with extremely high-load or very low load; this is shown in the high variability of the boxplot. On the contrary, when multiple users (i.e. 1:1024) are in the same PON, the aggregated offered traffic is more stable (i.e. less variability) and network designers may leverage from the statistical multiplexing properties to plan their PON networks. Although PONs with a very low split ratio are not economically feasible today, they have been included for the sake of theoretical completeness and to show the statistical multiplexing benefits of PONs with high split ratios.
\item The figure also reveals that current GPON technology can deal with those cases where the boxplots fall below the blue line limit of GPON. As shown, only the 1:1024 split case cannot be supported for 2016's traffic. However, it is worth remarking that the GPON standard allows at most 128 users. In addition, from a technical point of view, wide split PONs require a very high loss budget, which is a clear downside as well.
\end{itemize}

\begin{figure*}[!htbp]
\centering
\subfigure[Traffic needs, year 2025, 8.80~Mb/s peak per household]{
\includegraphics[width=0.48\textwidth]{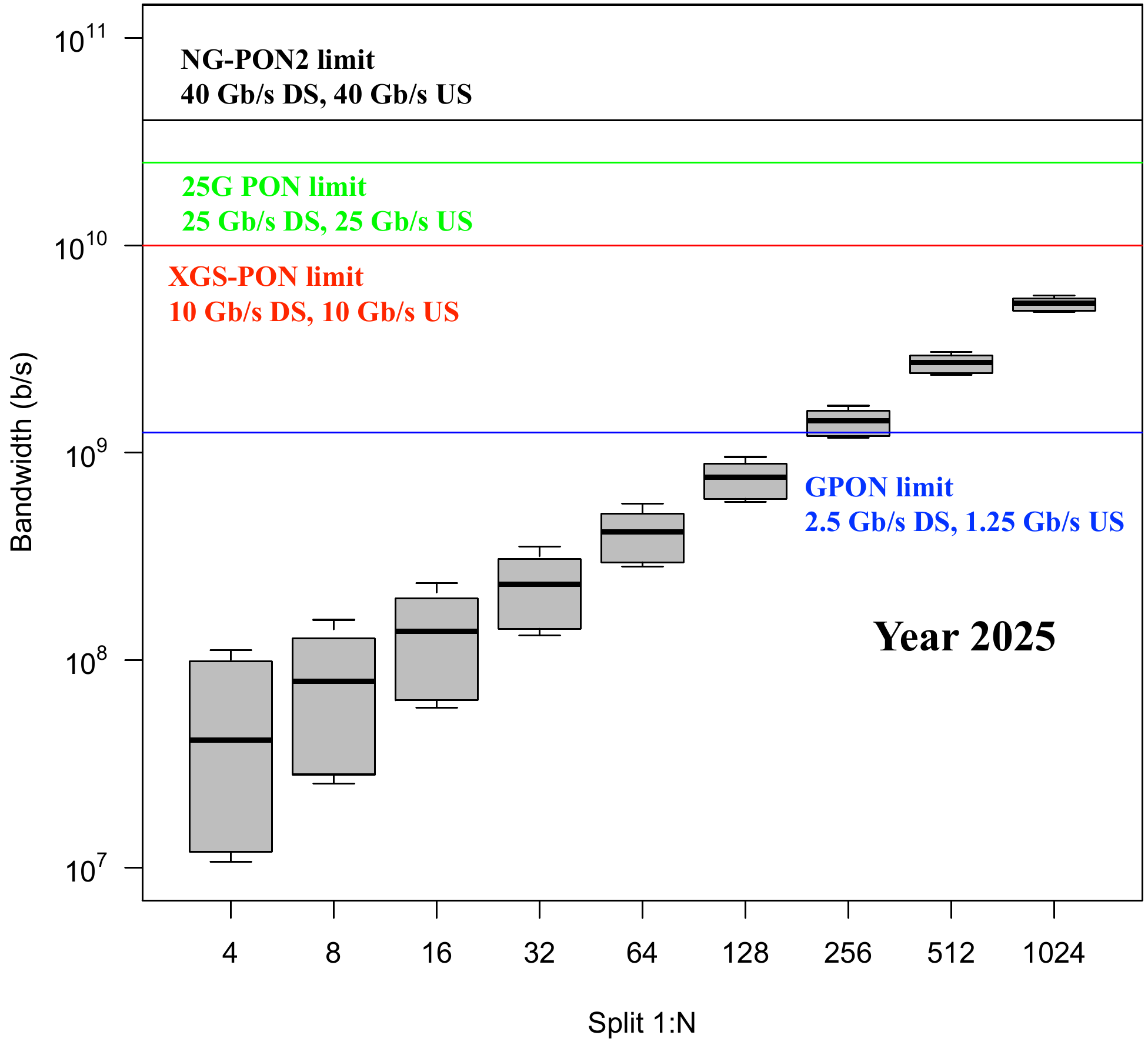}
}
\subfigure[Traffic needs, year 2030, 26.9~Mb/s peak per household]{
\includegraphics[width=0.48\textwidth]{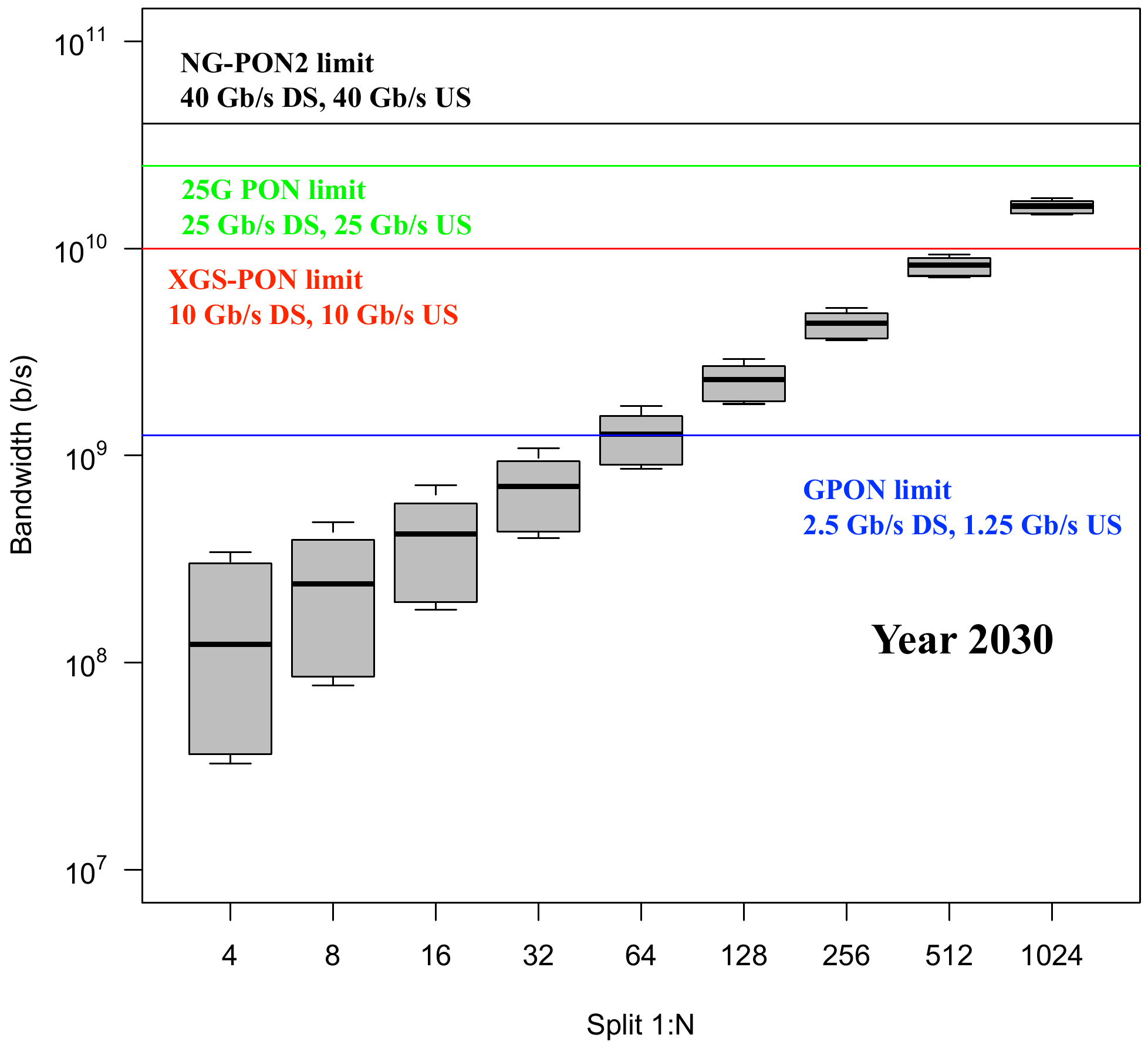}
}
\caption{Aggregated offered traffic per 1:N PON (Years 2025 and 2030)}
\label{figtrafficPONs}
\end{figure*}

Fig.~\ref{figtrafficPONs} shows the same experiments, this time for the traffic values forecasted for years 2025 and 2030, again along with the blue line (GPON limit at 1.25~Gb/s), red line (XGS-PON limit at 10~Gb/s), green line (25G PON limit at 25~Gb/s) and black line (NG-PON2 limit at 40~Gb/s). Again, the following observations are derived:

\begin{itemize}
\item With small split ratios (i.e. 1:16 and even 1:32), current GPON deployments may be sufficient up to year 2025. However, if larger split ratios are needed, the upgrade to XGS-PONs or 25G PONs will be necessary by 2030.
\item PONs with large split ratios like 1:256 and above have the benefit of statistical multiplexing properties, which reveals in the narrow boxplots in all figures. However, the rapid increase in traffic makes that such PONs serving so many residential users will definitely require an upgrade to PONs with larger capacity than 10G, i.e. 25G or 100G by 2030.
\end{itemize}

\section{PON upgrades: Possible evolution path}

\subsection{Reducing split or upgrading capacity}

\begin{figure}[!htbp]
\centering
\includegraphics[width=0.98\columnwidth]{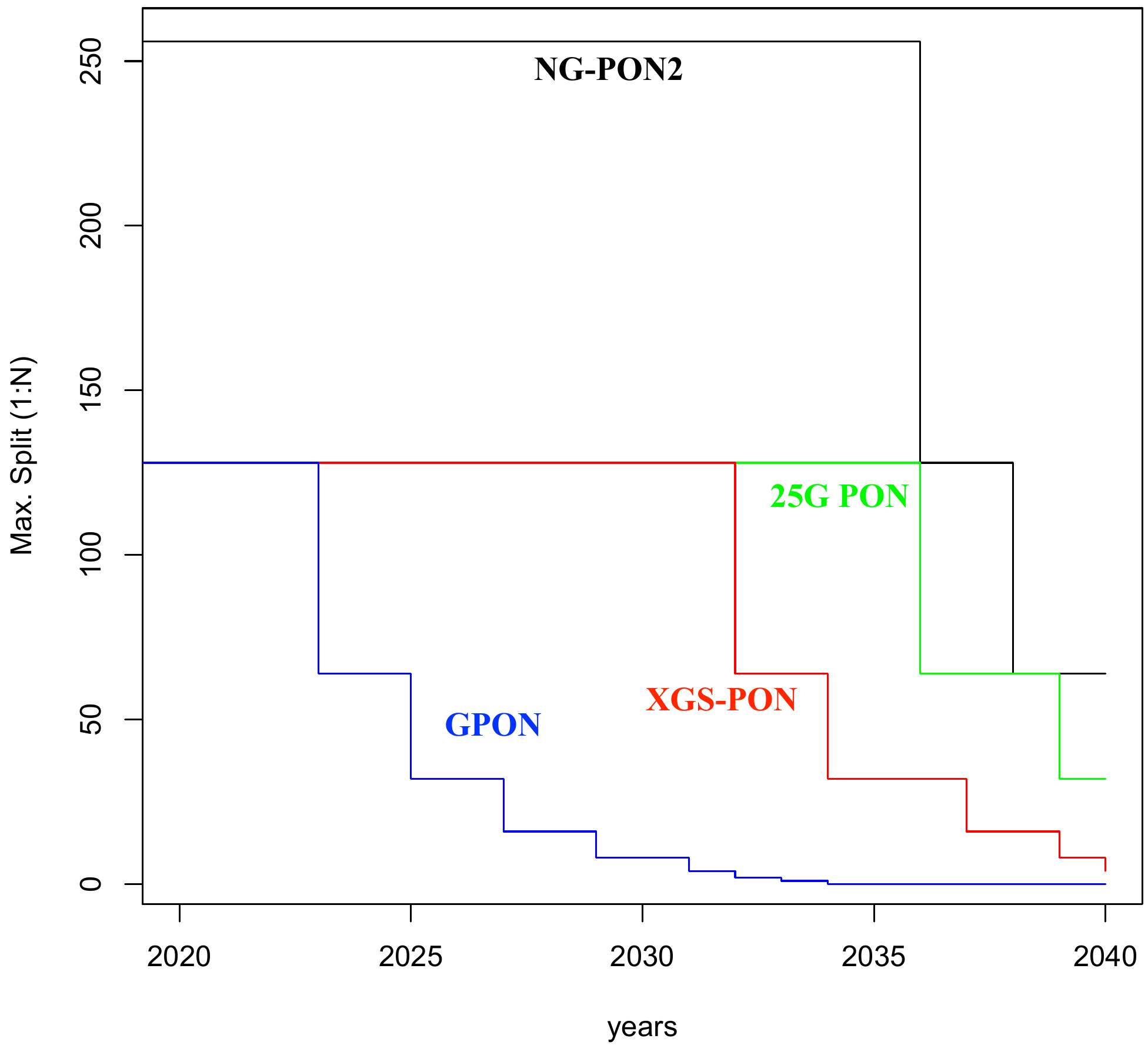}
\caption{Maximum split ratio per year and technology.}
\label{fig:evolutionpath}
\end{figure}

Fig.~\ref{fig:evolutionpath} shows the maximum split ratio for each PON technology that allows to meet the traffic needs of residential users in the next years. To find such values, we have simulated again the aggregated traffic needs of PON serving N residential users, then we check the 99-th percentile of traffic demands for such N users (after computing its 95\% confidence intervals with the Bootstrap method) and see if the 99-th percentile of the aggregated traffic demand falls below the 75\% percent limit of the upstream capacity offered by each PON standard. 

To illustrate this methodology, let us consider the following example: Consider the traffic needs at year 2025 for a GPON serving 64 users. Simulations show that such 64 users are expected to generate between 531 and 556~Mb/s (these are the 95\% Confidence Intervals for the median, i.e. 50-th percentile). However, confidence intervals for the 90-th percentile are between 754 and 799~Mb/s (i.e. 776$\pm$22.5~Mb/s), while the confidence intervals for the 99-th percentiles are between 961 and 1073~Mb/s (i.e. 1017$\pm$56~Mb/s). This means that, we are expected that 90\% of 1:64 GPONs deployed will generate accumulated traffic of about 776~Mb/s or less, while 99\% of the PONs deployed will generate traffic of about 1017~Mb/s or less (i.e. 1\% of the PONs will generate more traffic than 1017~Mb/s). Our criteria for upgrading a PON consists of checking that the 99-th percentile of aggregated traffic (i.e. 1017~Mb/s in this case) falls below 75\% of the upstream capacity of the PON, which is 937~Mb/s (i.e. 0.75x1.25~Gb/s). We can see that the condition is not met, therefore GPON with 1:64 is not a valid option in 2025; however, GPON with 1:32 is a valid scenario (its 99-th percentile is equal to 623~Mb/s, which falls below the limit of 937~Mb/s).

Following this methodology, we have computed the maximum split ratio allowed to satisfy the traffic demands for each year and PON technology. The following conclusions can be drawn from Fig.~\ref{fig:evolutionpath}:

\begin{itemize}
\item GPON with 1:64 split ratios and above will be a valid technology to satisfy the traffic demands of residential users until 2025. After this year, network operators will have to choose between decreasing the split ratio (to 1:32 or below) or upgrade to 10G XGS-PONs.  
\item XGS-PONs are expected to satisfy the traffic needs of up to 2035 with split ratios of 1:32 or above, and are expected to provide insufficient capacity by abour 2040. However, if large split ratios are needed (i.e 1:128), an upgrade to 25G PON or NG-PON2 will be necessary by about 2032.
\item Finally, 25G PON and NG-PON2 (i.e. 40G-TWDM-PON) are expected to provide enough bandwidth capacity to a large number of users for the next decades.
\end{itemize}

Concerning 40G-TWDM-PON and beyond, next targets in PON developments are expected to make lambdas go from 10 to 25~Gb/s and to enable 8 or even 16 lambdas per TWDM-PON, thus reaching a total aggregated capacity of 400G (i.e. 16x25G), see~\cite{Pon_roadmap}. Additional enhancements foreseen in~\cite{Pon_roadmap} envision that ONUs have access to more than one wavelength channel and expanding the optical bandwidth allowing for the transmission in both C+L bands. Another improvement is to consider the promising 25 Gb/s~\cite{pam4,pam4_v2} Pulse Amplitude Modulation with 4 levels (PAM-4) in PONs, an approach that has also been proposed by a number of manufacturers. However, the short reach supported by this technology suggests applicability only in dense-urban scenarios, where the distances between most users and the central office is below such distance limit.

Finally, an alternate evolution path to take into consideration is to use Wavelength Division Multiplexed PON (WDM-PON) technology\footnote{ITU-T Rec. G.698.3 Multichannel seeded DWDM applications with single-channel optical interfaces, approved in 2012}. In WDM-PON, a single wavelength is re-directed to an end user from the central office via a passive wavelength router (AWG) located in the outside plant.  This technology allows a single dedicated wavelength between each user and the central office, typically offering symmetric up- and down-stream bandwidth (either 1~Gb/s or even 10~Gb/s capacity). Typically, an AWG supports 32 ports, although there are wavelength routers available that can route up to 128 wavelengths. Advantages of WDM-PON include no bandwidth sharing between users, scalability, long-reach (given the low insertion-loss of filters, optional amplification), troubleshooting, security (users do not see other user 's traffic), and the possibility to individually adapt bitrates on a per-wavelength basis~\cite{rafa_commag}. Finally, ultra-dense WDM-PONs, featuring several hundreds of wavelengths operating at Gb/s or above have also been demonstrated in lab~\cite{coconut} and provide another possible technology to be taken into account in the future. However, it is worth noticing that replacing cost-effective passive splitters by active AWGs in the outside plant may not be economically feasible for network operators.

\subsection{Techno-economical aspects}

Upgrades of technology require investment from the network operators. The question here is whether it is more cost-effective to reduce the split ratio or upgrade ONU and OLT equipment to new standards. Essentially, any change in the outside plant is often very costly since ODNs account for 70\% of the total investments in PON deployments~\cite{huawei_whitepaper}. 

In this light, once the ODN is deployed, the split ratio may hardly ever be changed. This will depend on the operator’s strategy for splitter placement. The more centralised the splitter deployment is, the lower the cost of changing the split ratio. The extreme case of centralised splitting comprises having a single splitting level with all splitters located at the Central Office, so that no change is required in the outside plant.

However, in practice, many operators have decided to use a cascaded two-stage splitting architecture~\cite{schneir} in order to save fiber. In this setting, there is an initial (1:4 or 1:8) split stage in a closure not far from the central office or in the central office itself, and a second one (1:4 or 1:8) at the outside plant, yielding a total split between 1:16 and 1:64. In this case, the cost of a gradual reduction of just the first split stage can still become economically feasible, especially if those first-stage splitters were deployed in the central office. Obviously, split ratio reductions require more OLT ports and/or line cards.
  
Finally, a gradual reduction in the split ratio can be combined with capacity upgrades. Essentially, GPON and XGS-PON can coexist in the same ODN, since their wavelength plans do not overlap. This allows to smoothly migrate part of the users to XGS-PON (keeping others in GPON).

Hence, network operators have several degrees of freedom to upgrade their PONs and may design a migration strategy for the next decade, taking into account both technical and economic aspects of each PON technology.

\section{Summary and conclusions}

This article has overviewed the forecasted traffic needs of fixed access networks for the next decade, assuming current traffic profiles and expected CAGR of 25\%. Such traffic needs have been compared with current standard deployments to see when and how existing PON architectures will need to evolve, both in terms of users per PON and bandwidth capacity. We show that current Gigabit-PON deployments will result insufficient by 2030, requiring an upgrade to 10G-versions of PONs which, again will last for another 5 to 8 years approximately. Subsequent upgrades to 25G PON, NG-PON2 or above will be required after then. At that time, network operators will be required to consider not only the technical aspects of each PON technology available but also other economical aspects in their migration path.

In addition to this roadmap, this article shows the benefits of aggregating a large number of users under the same PON branch (namely 512 or even 1024) as proposed by LR-PONs, especially when the traffic profiles of users are highly uneven, as characterized by the Zipf distribution. However, a roadmap towards LR-PONs with large split ratios poses new challenges in terms of satisfying the bandwidth needs of so-many users, and are not standard yet, only scientific work demonstrated in lab. 

Finally, Passive Optical Networks are envisioned to provide broadband connectivity to not only residential users but also serve a mix of business and mobile end-points. In this light, such capacity upgrades will definitely need to occur before the previous dates since business traffic will surely be higher than residential along with mobile traffic, especially if the mobile industry goes towards the so-called Cloud Radio Access Networks (C-RAN) deployments where fronthaul traffic has strict latency and synchronization requirements along with bandwidth capacity needs. Future work shall investigate the impact of mixing such three types of traffic profiles, namely residential, business and mobile, and provide new PON evolution forecasts and migration path under such assumptions.

\section*{Acknowledgments}
The authors would like to acknowledge the support of the Spanish project TEXEO (grant no. TEC2016-80339-R), and the EU-funded project Fed4Fire (grant no. 318389) to the development of this work.

\ifCLASSOPTIONcaptionsoff
  \newpage
\fi

\section*{Biographies}

\begin{IEEEbiographynophoto}{Jos\'{e} Alberto Hern\'{a}ndez} (jahgutie@it.uc3m.es)
completed his five-year degree in Telecommunications Engineering at UC3M in 2002, and his Ph.D. degree in Computer Science at Loughborough University, Leicester, United Kingdom, in 2005. He has been a senior lecturer in the Department of Telematics Engineering since 2010, where he combines teaching and research in the areas of optical WDM networks, next-generation access networks, metro Ethernet, energy efficiency, and hybrid optical-wireless technologies. He has published more than 100 articles in both journals and conference proceedings on these topics. 
\end{IEEEbiographynophoto}

\begin{IEEEbiographynophoto}{Rafael S\'{a}nchez} (rsfuente@it.uc3m.es) 
holds an Ph.D. in Telematics Engineering (2016) by University Carlos III de Madrid (UC3M) and a Degree in Telecommunications Engineering (1996) by Polytechnic University of Valencia. Since 1996, he has been involved in multiple enterprise projects in areas like optical networks (SDH/DWDM), IPTV, Ethernet fiber access and networking in companies like Lucent Technologies and Nortel. Currently, he works for Google as Cloud Architect.
\end{IEEEbiographynophoto}

\begin{IEEEbiographynophoto}{Ignacio Mart\'{i}n} (ignmarti@it.uc3m.es) 
completed the BEng. in Telematics Engineering and the MSc. Cybersecurity at Universidad Carlos III de Madrid in 2014 and 2015 respectively. At present, he is pursuing the Ph.D. degree at Universidad Carlos III de Madrid in the areas of Data Analytics applied to computer security and networking.
\end{IEEEbiographynophoto}

\begin{IEEEbiographynophoto}{David Larrabeiti} (dlarra@it.uc3m.es)
is with the Telematics Engineering Department of University Carlos III of Madrid (UC3M) since 1998, where he is professor of Switching and Optical Networks. He is currently involved in a number of EU research projects on new optical networking paradigms, including PASSION, BlueSPACE and Metro-haul. He has also served in the TPC of ECOC, HPSR, Globecom and other conferences. His current research interests include all-optical ultra-low latency switching and tactile Internet.
\end{IEEEbiographynophoto}

\balance

\end{document}